\documentclass[aps,amsfonts,preprint,nofootinbib]{revtex4}
\def\S2{\bar{S}}
\def\a{\alpha}
\def\b{\beta}

\def\t{\tau}
\def\an{a_{n}}
\def\ta{\tilde{a}_n}
\def\and{a_{n}^\dagger}

\def\tab{\tilde{\bar{a}}_n}

\def\sn2d{\Sn2^\dagger}

\def\({\left(}
\def\){\right)}
\def\<{\left\langle}
\def\>{\right\rangle}
\def\s{\sigma}
\begin{document}

\title{Closed String Thermal Torus From Thermo Field Dynamics}

\author{M. C. B Abdalla, A. L. Gadelha and Daniel L. Nedel}

\affiliation{Instituto de F\'{\i}sica Te\'{o}rica, Unesp,
Pamplona 145, S\~{a}o Paulo, SP, 01405-900, Brazil }

\begin{abstract}
In this Letter a topological interpretation for the string
thermal vacuum in the Thermo Field Dynamics (TFD) approach
is given. As a consequence, the relationship between the
Imaginary Time and TFD formalisms is achieved when both are
used to study closed strings at finite temperature.
The TFD approach starts by duplicating the system's degrees of
freedom, defining an auxiliary (tilde) string. In order to
lead the system to finite temperature a Bogoliubov
transformation is implemented. We show that the effect of
this transformation is to glue together the string
and  the tilde string  to obtain a torus. 
The thermal vacuum appears as the boundary state
for this identification. Also, from the thermal state
condition, a Kubo-Martin-Schwinger condition for the torus
topology is derived.
\end{abstract}

\maketitle

\section{Introduction}

Over the years string theory has been considered the best
candidate for a theory that quantizes gravity. The two
most pressing areas are those of black holes physics and
cosmology.
However, in despite of the significant progress which
has been made in the last years, there is a lot of
indications that to understand the non-perturbative
regime  of the theory (necessary to study black holes
and cosmology), a more robust theoretical framework is
needed. The formulation of string theory at finite
temperature has  given good hints that the perturbative
formulation of the theory does not work at high
temperature. Owing to the exponential growth of states
as function of energy, there is an upper bound temperature
above which the statistical partition function diverges (the
Hagedorn temperature). The very existence of the Hagedorn
temperature shows that the fundamental degrees of freedom
of the theory could not be the ones of the perturbative
string. If the specific heat at the Hagedorn temperature
is finite it denotes a phase transition, and just as quark
and gluons emerge as the basic ingredients of QCD at high
temperature, the true degrees of freedom of the string
theory may also emerge at high temperature. So, beyond
phenomenological applications, string at finite
temperature may also provide important pieces of evidence
of the true degrees of freedom of the theory at
non-perturbative regime.

It has to be mentioned that in the last two decades there
have been interesting works studying string theory at finite
temperature. The standard way used to lead the string
theory from zero to finite temperature is the Imaginary
Time formalism \cite{mat1,le1,adas}.

In general, the statistical average of an operator
$Q$, $\left\langle  Q \right\rangle$, is defined by
the  functional $ \omega\(Q\)=Tr\(\rho Q\)$:
\begin{equation}
\left\langle  Q \right\rangle
= \frac{{\mbox Tr}[Q e^{-\b H}]}{\omega\(1\)},
\end{equation}
where $\rho = e^{-\b H}$ and $\omega\(1\)$ is the partition
function.  In statistical quantum field theory the trace
is taken over the Fock space formed by the field
operators equipped with some algebraic properties. As a
consequence of the cyclic property of the trace, the
Kubo-Martin-Schwinger condition follows:
$\omega\(A\(t\)B\)=\omega\(BA(t+i\b)\)$, which is the basis of
the Imaginary Time formalism (Matsubara formalism).

Using the Matsubara formalism, a quantum field at finite
temperature relates the functional generator in $R^{1,d-1}$ to a
trace when the theory is defined in $R^{d-1}\times S^1$ with the
length of the compactified circle equals to $\b$. When the Matsubara
formalism is applied to a free closed string, the theory is
defined on a torus, where $\b$ is related to the imaginary
part of the modulus parameter $\tau$.  
The real part of $\t$ is related to a Lagrange multiplier
$(\lambda)$, that imposes the $S^1$ isometry of the closed
string to the Fock space. The gauge fixing of the $S^1$ isometry
improves the level matching condition of the physical spectrum of
the closed string. So, for the closed string, $ \omega\(Q\)$ must
be redefined as:
\begin{equation}
\omega\(Q\)= \int_{0}^{1} d\lambda Tr\left[Q
e^{-\b H+2\pi i\lambda P}\right],
\label{tr2}
\end{equation}
where $P$ is the momentum operator that generates translations in
the world-sheet $\sigma$ coordinate. The integral over the Lagrange
multiplier guarantees that the trace is taken only over the
physical states. The partition function is defined on a  torus
with moduli space parameters defined by:
$\t=\lambda+i\frac{\b}{2\pi}$. 
A different perspective in which 
the torus topology can be observed is perceived by noting that
the operator $e^{\(-2\pi\beta H\)}$
propagates a closed string through imaginary time $i2\pi\b$ and
the operator $e^{i2\pi \lambda P}$ rotates the closed string at an
$2\pi\lambda$ angle. Hence the trace corresponds to a torus
constructed by gluing together the ends of an open cylinder with a
relative twist \cite{laz}.

On the other hand, the functional $\omega\(Q\)$ is called a
state in algebraic statistical quantum field theory and
the operators form a $C^{\star}$ algebra \cite{haag,emch}.
The functional $\omega$ resembles a vector space, so that the
algebra equipped with a particular functional admits a
reducible representation on a Hilbert space such as a Fock
space. This is the basis of an alternative formalism to study
quantum field at finite temperature developed by Takahashi
and Umezawa, named Thermo Field Dynamics (TFD)
\cite{ume2,ume4,rev2,ume1,kha2,kha3}. The TFD was developed
in order to handle finite temperature with a real time
operator formalism \cite{kob1,leb1}. The main idea is to
interpret the statistical average  as the expectation value
of $Q$ in a thermal vacuum:
\begin{equation}
\frac{ \omega\(Q\)}{\omega\(1\)}= \left\langle 0 (\b ) \left| Q
\right| 0 (\b )\right\rangle .
\label{tr1}
\end{equation}

Concerning string theory, the idea of building a thermal Fock
space is particularly tempting as new degrees of freedom
could be identified in this Fock space at some temperature.
Although the TFD approach was adopted in the past to study first
quantized bosonic string \cite{FNA1,FNA2}, heterotic string
\cite{YLE3,FNA3} and string field theory \cite{YLE1}, it was
employed within a path integral formulation. The idea of using the
Fock space formulation in string theory came up in Refs.
\cite{IVV,AGV2,AEG,AGV3,AGV4,AGV5}, where the thermal space was
used to construct bosonic thermal boundary states interpreted as
D-branes at finite temperature.
To further explore the algebraic characteristics of the TFD, and
to upgrade it to a powerful tool to understand string theory at
finite temperature, it is first necessary to set up the
connections between the TFD and the Imaginary Time formalisms,
when both are applied to strings at thermal equilibrium.
In \cite{nos}, the TFD was used to derive thermodynamical
quantities for type IIB superstring in a pp-wave background.
It was shown that the free energy, calculated from the 
world-sheet torus partition function in the Imaginary
Time formalism, can be derived from the thermal expectation
value in TFD.
In  subsequent works the $SU(1,1)$ and $SU(2)$ thermal groups
were used to generalize the TFD for applications in closed string
theories \cite{nos2,nos3}. Such a generalization again reproduces
the free energy obtained via Imaginary Time formalism. A question
arises: how the torus defined by taking the trace on the right-hand
side of equation $\(\ref{tr1}\)$ can be interpreted on the left-hand
side?  The aim of this work is to answer this question. It will be
shown how a torus can appear when the TFD is used to treat the
closed string at finite temperature at
thermal equilibrium\footnote{The closed bosonic string was chosen
for the sake of simplicity, but the results presented here can
easily be extended for the light-cone Green-Schwarz superstring.}.
Also, it will be shown that the thermal vacuum for the free closed
string represents a kind of boundary state for string theory.
As a consequence, a Kubo-Martin-Schwinger (KMS) relation arises
with a topological interpretation.

\section{TFD approach for the closed bosonic string}

The TFD is introduced by first duplicating the degrees of
freedom of the system. To this end a copy of the original
Hilbert space, denoted by $\widetilde{H}$, is constructed.
The tilde Hilbert space for the closed bosonic string in
the light-cone gauge is built with a set of oscillators
that have the same commutation properties as the original
ones. The total Hilbert space is the tensor product of the
two spaces
${\cal H} _{T}={\cal H}\otimes\widetilde{{\cal H}}$
with elements
$|\Phi \rangle =|\phi,\widetilde{\phi }\rangle $
and the vacuum is defined by
\begin{eqnarray}
{a}^{I}_{n}|0 \rangle&=&{\bar{a}_{n}}^{I}|0 \rangle = 0,
\nonumber
\\
\ta^I|0 \rangle&=&\tab^I|0 \rangle = 0,
\end{eqnarray}
for $n>0$ and $|0 \rangle
=\left.\left|0\right\rangle \!\right\rangle
=|0\rangle \otimes|\widetilde{0} \rangle$ as usual.
The bar and non-bar operators denote, here and in the
following, the left- right-moving modes, respectively.
The map between the tilde and non-tilde operators is
defined by the following tilde (or dual)
conjugation rules \cite{kha5}:
\begin{eqnarray}
(A_{i}A_{j})\widetilde{}
&=&\widetilde{A}_{i}\widetilde{A}_{j}, \nonumber \\
(cA_{i}+A_{j})\widetilde{} &=&c^{\ast
}\widetilde{A}_{i}+\widetilde{A}_{j}, \nonumber \\
(A_{i}^{\dagger })\widetilde{},
&=&(\widetilde{A}_{i})^{\dagger }, \nonumber \\
(\widetilde{A}_{i})\widetilde{} &=&A_{i},  \nonumber \\
\lbrack \widetilde{A}_{i},A_{j}] &=&0.  \label{til}
\end{eqnarray}
As the operators of the two systems commute among
themselves, the doubled system is described by two
independent strings defining two world-sheets.
Using the Euclidian time $\(\tau=-it\)$, the  mode
expansions for the two strings in the light-cone gauge are:
\begin{eqnarray}
X^I = x^I_0 - i\a^\prime p^I_0 t
+\sqrt{\frac{\a^\prime}{2}}\sum_{n > 0}\frac{1}{\sqrt{n}}
\bigg[\(\an^I e^{-n\(t-i\s \)}
+ \an^{\dagger \: I}e^{n\(t -i\s\)}\)
\nonumber
\\
+ \({\bar a}_{n}^{I} e^{-n(t+i\s)}
+{\bar a}_{n}^{\dagger \: I}
e^{n(t + i\s)}\)\bigg],
\end{eqnarray}
and
\begin{eqnarray}
\widetilde{X}^I =
\widetilde{x}^I_0 + i\a^\prime \widetilde{p}^I_0 t
+\sqrt{\frac{\a^\prime}{2}}\sum_{n > 0}\frac{1}{\sqrt{n}}
\bigg[\(\ta^I e^{-n\(\tilde{t}+i\tilde{\sigma}\)}
+\ta^{\dagger \: I}e^{n\(\tilde{t}+i\tilde{\sigma}\)}\)
\nonumber
\\
+\(\tab^{I}e^{-n\(\tilde{t}-i\tilde{\sigma}\)}
+\tab^{\dagger\: I}e^{n\(\tilde{t}-i\tilde{\sigma}\)}\)\bigg],
\end{eqnarray}
where the expansion for $\widetilde{X}\(t,\sigma\)$ was
obtained from the tilde conjugation rules. It can be seen
that the tilde string can be interpreted as a string
propagating backwards in Euclidian time.

The oscillators satisfy the extended algebra:
\begin{eqnarray}
\left[a_{n}^I,a_{m}^{\dagger \:J}\right]
&=&\left[\tilde{a}^{I}_{n},\tilde{a}_{m}^{\dagger\:J}\right]
=\delta_{n,m}\delta^{I,J},
\nonumber
\\
\left[\bar{a}_{n}^{I},\bar{a}_{m}^{\dagger\:J}\right]
&=&\left[\tilde{\bar{a}}^{I}_{n},
\tilde{\bar{a}}_{m}^{\dagger\:J }\right]
=\delta_{n,m}\delta^{I,J},
\nonumber
\\
\left[a_{n}^{\dagger\:I},\tilde{a}_{m}^{J}\right]
&=&\left[a_{n}^{\dagger\:I},\tilde{a}_{m}^{\dagger\:J}\right]
=\left[a_{n}^{I},\tilde{a}_{m}^{J}\right]=
\left[a_{n}^{I},\tilde{a}_{m}^{\dagger\:J}\right]=0.
\end{eqnarray}

Now, the thermal vacuum and the thermal Fock space
for the closed string can be constructed. This is achieved
by implementing a Bogoliubov transformation in the total
Hilbert space. At this point it is necessary to emphasize
that the physical variables are described by the non-tilde
operators. The tilde operators are auxiliary degrees
of freedom necessary to provide enough room to accommodate
the thermal properties of the system. All the thermodynamical
quantities are derived from expectation values of $T=0$
non-tilde operators in the thermal vacuum to be defined.

The generator for the Bogoliubov transformation of the closed
string is given by
\begin{equation}
G=G+\bar{G},
\end{equation}
where
\begin{eqnarray}
G &=&-i\sum_{n}\theta_{n}\(a_{n}\cdot {\tilde a}_{n}-{\tilde
a}_{n}^{\dagger} \cdot a_{n}^{\dagger}\),
\nonumber
\\
{\bar G}&=&-i \sum_{n}{\bar \theta}_{n}\({\bar a}_{n}\cdot{\tilde
{\bar a}}_{n}-{\tilde {\bar a}}_{n}^{\dagger} \cdot {\bar
a}_{n}^{\dagger}\). \label{gen}
\end{eqnarray}
Here the dots represent the inner products and
$\theta$, ${\bar \theta}$ are the tranformation's parameters.
As we shall see, at thermal equilibrium they are related to
the Bose-Einstein distribution of the oscillator
$n$.

The thermal vacuum is given by the following relation
\begin{eqnarray}
\left |0\(\theta\)\right\rangle &=& e^{-i{G}}\left.
\left|0\right\rangle \!\right\rangle \nonumber
\\
&=&
\prod_{n=1}\left[\left(
\frac{1}{\cosh(\theta_{n})}\right)^{D-2}\left( \frac{1}{\cosh({\bar
\theta}_{n})}\right)^{D-2}
e^{\tanh\(\theta_{n}\)\(a_{n}^{\dagger}\cdot {\tilde
a}_{n}^{\dagger}\)+ \tanh\({\bar \theta}_{n}\)\({\bar
a}_{n}^{\dagger}\cdot {\tilde {\bar a}}_{n}^{\dagger}\)}\right ]\left.
\left|0\right\rangle\!\right\rangle
\label{tva}.
\end{eqnarray}

The creation and annihilation operators at $T \neq 0$ are given by
the Bogoliubov transformations as follows
\begin{eqnarray}
a_{n}^{I}\(\theta_{n}\)&=&e^{-iG}a_{n}^{I}e^{iG}
=\cosh\(\theta_{n}\)a_{n}^{I} -
\sinh\(\theta\){\widetilde a}_{n}^{\dagger \: I},
\\
{\bar a}_{n}^{I}\({\bar \theta}_{n}\)&=&e^{-iG}{\bar
a}_{n}^{I}e^{iG} =\cosh\({\bar \theta}_{n}\){\bar a}_{n}^{I} -
\sinh\(\bar{\theta}_{n}\){\widetilde {\bar a}})_{n}^{\dagger \: I}.
\end{eqnarray}
These operators annihilate the state written in $\(\ref{tva}\)$ defining
it as the vacuum. By using the Bogoliubov transformation, the relations
\begin{eqnarray}
a_{n}^{I}\(\theta_{n}\)\left |0\(\theta\)\right\rangle &=& \widetilde
{a}_{n}^{I}\(\theta_{n}\)\left |0\(\theta\)\right\rangle = 0,
\nonumber
\\
\left\langle 0\(\theta\)\right| a_{n}^{ \dagger \: I}\(\theta_{n}\)
&=&\left\langle 0\(\theta\)\right|\widetilde{a}_{n}^{ \dagger \: I}
\(\theta_{n}\)= 0,
\end{eqnarray}
give rise the so called thermal state conditions:
\begin{eqnarray}
\left[a_{n}^{I}-
\tanh\(\theta_n\)\widetilde{a}^{\dagger \: I}_n\right]
\left|0\(\theta\)\right\rangle
&=&0,
\label{cond1}
\\
\left[\widetilde{a}_{n}^{I}-
\tanh\(\theta_n\){a}^{\dagger\: I}_{n}\right]
\left|0\(\theta\)\right\rangle
&=&0,
\label{cond2}
\\
\left[\bar{a}_{n}^{I}
-\tanh\(\bar{\theta}_n\)\widetilde{\bar{a}}^{\dagger\: I}_n\right]
\left|0\(\theta\)\right\rangle
&=&0,
\label{cond3}
\\
\left[\widetilde{\bar{a}_{n}^{I}}
-\tanh\(\bar{\theta}_n\) {\bar{a}}^{\dagger\: I}_{n}\right]
\left|0\(\theta\)\right\rangle
&=&0,
\label{cond4}
\\
\left\langle 0\(\theta\)\right|\left[a^{\dagger\: I}_{n}-
\tanh\(\theta_n\)\widetilde{a}_{n}^{I}\right]
&=&0,
\label{cond5}
\\
\left\langle 0\(\theta\)\right|\left[\widetilde{a}^{\dagger\: I}_{n}
-\tanh\(\theta_n\)a_{n}^{n}\right]
&=&0,
\label{cond6}
\\
\left\langle 0\(\theta\)\right|\left[\bar{a}^{\dagger\: I}_{n}
-\tanh\(\bar{\theta}_n\)\widetilde{\bar{a}}_{n}^{I}\right]
&=&0,
\label{cond7}
\\
\left\langle 0\(\theta\)\right|\left[\widetilde{\bar{a}}^{\dagger\: I}_{n}
-\tanh\(\bar{\theta}_n\)\bar{a}_{n}^{I}\right]
&=&0.
\label{cond8}
\end{eqnarray}

The thermal Fock space is constructed by applying the
thermal creation operators to the vacuum $\(\ref{tva}\)$.
As the Bogoliubov transformation is canonical, the thermal
operators obey the same commutation relations as the operators
at $T=0$. It is easy to see that thermal states are not
eigenstates of the original Hamiltonian but they are
eigenstates of the combination:
\begin{equation}
{\widehat H} = H -{\widetilde H},
\label{hath}
\end{equation}
in such a way that ${\widehat H}$ plays the r\^{o}le of the
Hamiltonian, generating  temporal translation in the thermal
Fock space. Using the commutation relations we can prove that
the Heisenberg equations are satisfied replacing $H$ and
${\widetilde H}$ by ${\widehat H}$. Also we have
\begin{equation}
\widehat{P}= P-\widetilde{P},
\label{hath2}
\end{equation}
where $P$, $\widetilde{P}$ and ${\widehat P}$ are the world-sheet
translation generators of the original, auxiliary and transformed
systems, respectively.

The effect of the Bogoliubov transformation is to entangle the
elements of the two Hilbert spaces. After the transformation,
the image of the two independent strings defining two different
cylinders is lost. In quantum field theory the thermal vacuum
can be interpreted as a condensed state and the Bogoliubov
transformation confines the fields in a restricted region of
the time axis. In \cite{ademir} the analytical continuation
for $\sinh^{2}\(\theta_{n}\)$ is explored to demonstrate that
the Bogoliubov operator can also produce confinement in spatial
directions. In the next section, it will be demonstrated that
the time confinement produced by the Bogoliubov operator, in a
very special way, transforms the two cylinders in a torus and
the state $\(\ref{tva}\)$ becomes a string boundary state.

\section{Thermal Torus and Thermal State}

Suppose that one wants to build a torus with the two initial
cylinders defined by two independent strings. From the tilde
conjugation rules one sees that the tilde string propagates
backwards in the Euclidian time. In this way, considering the
original string propagating by an amount of the value
$\frac{\beta}{2}$, i.e., $X\(\frac{\beta}{2}\)$, then the tilde
string propagates by the same amount in the opposite
direction, $\widetilde{X}\(-\frac{\beta}{2}\)$. One can construct
a torus by gluing together the end of the original cylinder with
the origin of the tilde cylinder, and vice-versa. This procedure will
confine the fields to a restrict region $\beta$ in the Euclidian
time. Also, before gluing, the identification
$\tilde{\sigma}=\sigma-\pi\lambda$, must be done in order to take
into account the Dehn twist in one cycle. The two parameters of
the resulting torus moduli space will be related to $\b$ and
$\lambda$. The above considerations can be written as follows:
\begin{eqnarray}
X\(t,\sigma\)
-\widetilde{X}\(-t-\frac{\b}{2},\sigma-\lambda\pi\)&=&0,
\nonumber
\\
X\(-\tilde{t}-\frac{\beta}{2},\tilde{\sigma}+\lambda\pi\)
-\widetilde{X}\(\tilde{t},\tilde{\sigma}\)&=&0.
\label{ident}
\end{eqnarray}
Expanding $X\(t,\sigma\)$ and $\widetilde{X}\(\widetilde
{t},\widetilde{\sigma}\)$ in  modes, the above identification
turns out to be the following set of operatorial equations for a
boundary state $|\Phi \rangle =|\phi,\widetilde{\phi }\rangle $:
\begin{eqnarray}
\left[a_{n}^{I}-
e^{-n\(\frac{\b}{2}+i\lambda\pi\)}\widetilde{a}^{\dagger\: I}_{n}
\right]|\Phi \rangle
&=&0,
\label{bond1}
\\
\left[\widetilde{a}_{n}^{I}-
e^{-n\(\frac{\b}{2}+i\lambda \pi\)}{a}^{\dagger\: I}_{n}\right]
|\Phi \rangle &=&0,
\label{bond2}
\\
\left[\bar{a}_{n}^{I}
-e^{-n\(\frac{\b}{2}-i\lambda \pi\)}
\widetilde{\bar{a}}^{\dagger\: I}_{n}\right]
|\Phi \rangle
&=&0,
\label{bond3}
\\
\left[\widetilde{\bar{a}}_{n}^{I}
-e^{-n\(\frac{\b}{2}-i\lambda \pi\)}{\bar{a}}^{\dagger\: I}_{n}
\right]|\Phi \rangle &=&0.
\label{bond4}
\end{eqnarray}
It is assumed that there is no center-of-mass momentum
dependence on $|\Phi \rangle$, which means that the
theory is being described in a particular frame. As the
state $|\Phi \rangle$ will be related to the thermal
state, there is no problem with that assumption since the
temperature breaks the Lorentz invariance. Defining
$\t= \lambda+i\frac{\b}{2\pi}$ and $q=e^{2\pi i \t}$,
the normalized solution of the equations
$\(\ref{bond1}\)$-$\(\ref{bond4}\)$ is
\begin{equation}
|\Phi \rangle=
\left[\(q\bar{q}\)^\frac{D-2}{24}|\eta(\t)|^{-2\(D-2\)}\right]
^{-\frac{1}{2}}
e^{\sum_{n>0}
\(a_{n}^{\dagger}\cdot {\tilde a}_{n}^{\dagger}\)q^\frac{n}{2}}
\times
e^{\sum_{n>0}
\bar{a}_{n}^{\dagger}\cdot {\tilde {\bar a}}_{n}^{\dagger}\bar{q}
^\frac{n}{2}}\left.\left|0\right\rangle\!\right\rangle,
\label{bounds}
\end{equation}
where $\eta\(\t\)$ is the Dedekind $\eta$ function
\begin{equation}
\eta\(\t\)=q^{\frac{1}{24}}\prod_{n=1}\(1-q^n\),
\end{equation}
and $\bar{q}$ is the complex conjugate of $q$.

Next it will be shown that the above state is precisely
the thermal state $\(\ref{tva}\)$. To this end it is
necessary to find the explicit dependence of the
$\theta$  parameters with respect to $\beta$ and
$\lambda$. If the following identifications
\begin{equation}
\tanh\(\theta_n\)=\bar{q}^{\frac{n}{2}},
\qquad
\tanh\(\bar{\theta}_n\)=q^{\frac{n}{2}},
\label{tan}
\end{equation}
hold, one can see that equations
$\(\ref{bond1}\)$-$\(\ref{bond4}\)$
are in fact the thermal state conditions
$\(\ref{cond1}\)$-$\(\ref{cond4}\)$ that define
the thermal vacuum. Equation $\(\ref{tan}\)$ can be derived by
minimizing the following potential $F$ \cite{ume2,nos,nos2}:
\begin{equation}
F=E-\frac{1}{\beta }S,
\label{f}
\end{equation}
with respect to the transformation's parameters.
Here $E$ is related with the thermal energy and $S$ with the
entropy of the string. So, $F$ is a free energy like potential.
In TFD, the thermal energy is given by computing the matrix
elements of the $T=0$ Hamiltonian in the thermal vacuum.
In the same way, $S$ is computed by using the entropy operator
$K$ defined for the closed string by:
\begin{eqnarray}
K &=&
-\sum_{n=1}
\bigg\{ a_{n}^{\dagger }\cdot a_{n}
\ln \left( \sinh^{2}\left(\theta _{n}\right)\right)
- a_{n}\cdot a_{n}^{\dagger }
\ln \left( \cosh^{2}\left(\theta _{n}\right)\right)\bigg\}
\nonumber
\\
&-&\sum_{n=1}\left\{ \bar{a}_{n}^{\dagger }\cdot\bar{a}_{n}\ln
\left( \sinh^{2}\left( \bar{\theta }_{n}\right)\right)
-\bar{a}_{n}\cdot \bar{a}_{n}^{\dagger }
\ln \left( \cosh^{2}\left(\bar{\theta }_{n}\right)\right)\right\}.
\label{k}
\end{eqnarray}
 
In order to apply TFD in closed strings, it is necessary
to redefine the thermal energy operator in such a way that
the level matching conditions are improved. The procedure
is to consider as the thermal energy operator, the
shifted Hamiltonian in the sense of Refs. \cite{OJI,nos,nos2},
\begin{equation}
H_{s} = \frac{P^{i} \cdot P^{i}}{2}+
\sum_{n=1}n\left( N_{n}+\bar{N}_{n}\right)
+\frac{1}{\beta}i2\pi\lambda
\sum_{n=1}n\left( N_{n}-\bar{N}_{n}\right)-2,
\label{hshif}
\end{equation}
where
\begin{equation}
N_{n}= a_{n}^{\dagger}\cdot a_{n},
\qquad
\bar{N}_{n}= \bar{a}_{n}^{\dagger}\cdot \bar{a}_{n},
\label{hs}
\end{equation}
and $\sum_{n=1}n\left( N_{n}-\bar{N}_{n}\right)$
is the momentum operator of the world-sheet.
By considering the expectation value of that
shifted Hamiltonian as the thermal energy, the $S^1$
isometry is fixed and the expectation value $\(\ref{tr1}\)$
reproduces the trace $\(\ref{tr2}\)$ \cite{nos,nos2}.

Now, using $\(\ref{hs}\)$ and $\(\ref{k}\)$ $F$ is
minimized and the following relations are found:
\begin{equation}
\sinh^{2}\(\theta_{n}\)
=\frac{1}{e^{n\left(\beta+i2\pi\lambda \right)}-1},
\qquad
\sinh^{2}\( \bar{\theta}_{n}\)
=\frac{1}{e^{n\left(\beta-i2\pi\lambda\right)}-1},
\end{equation}
These expressions fix the thermal vacuum $\(\ref{tva}\)$,
now reproducing the trace over the transverse sector.
All the thermodynamical quantities at equilibrium can be derived
from this state. However, a time dependence could be allowed on
$\theta$ and out of equilibrium physics could be described.
In this sense, this formalism is more general than the
Imaginary Time formalism. 
Also, we can easily see that
\begin{equation}
\prod_{n=1}\left(
\frac{1}{\cosh(\theta_{n})}\right)^{D-2}
\left( \frac{1}{\cosh({\bar\theta}_{n})}\right)^{D-2}
=\left[\(q\bar{q}\)^\frac{D-2}{24}|\eta(\t)|^{-2\(D-2\)}
\right ]^{-\frac{1}{2}},
\end{equation}
so, equations $\(\ref{cond1}\)$-$\(\ref{cond8}\)$
hold and the thermal vacuum is exactly the boundary
state given in $\(\ref{bounds}\)$.

\section{KMS condition for the closed bosonic string}

In this section the thermal state (boudary) equations
will be used in order to derive a KMS condiction for
closed bosonic string.

The operators $\hat{H}$ and $\hat{P}$ defined in
$\(\ref{hath}\)$ and $\(\ref{hath2}\)$ commute with the
Bogoliubov generator $\(\ref{gen}\)$. Therefore, it follows
that
\begin{equation}
\hat{H}\left |0\(\theta\)\right\rangle
= \hat{P}\left |0\(\theta\)\right\rangle = 0.
\end{equation}
Now, using the following commutation relations
\begin{eqnarray}
e^{\(\frac{\beta}{2}\hat{H}+i\pi\lambda\hat{P}\)}
a_n^{I}
e^{-\(\frac{\beta}{2}\hat{H}+i\pi\lambda\hat{P}\)}
&=&e^{-n\(\frac{\b}{2}+i\lambda\pi\)}a_n^{I},
\nonumber
\\
e^{\(\frac{\beta}{2}\hat{H}+i\pi\lambda\hat{P}\)}
a_n^{\dagger\:I}
e^{-\(\frac{\beta}{2}\hat{H}+i\pi\lambda\hat{P}\)}
&=&e^{n\(\frac{\b}{2}+i\lambda\pi\)}a_{n}^{\dagger\:I},
\nonumber
\\
e^{\(\frac{\beta}{2}\hat{H}+i\pi\lambda\hat{P}\)}
\widetilde{a}_n^{I}
e^{-\(\frac{\beta}{2}\hat{H}+i\pi\lambda\hat{P}\)}
&=&e^{n\(\frac{\b}{2}+i\lambda\pi\)}\widetilde{a}_n^{I},
\nonumber
\\
e^{\(\frac{\beta}{2}\hat{H}+i\pi\lambda\hat{P}\)}
\widetilde{a}_n^{\dagger\:I}
e^{-\(\frac{\beta}{2}\hat{H}+i\pi\lambda\hat{P}\)}
&=&e^{-n\(\frac{\b}{2}+i\lambda\pi\)}
\widetilde{a}_n^{\dagger\:I},
\label{comut}
\end{eqnarray}
and $\(\ref{tan}\)$, the equations $\(\ref{cond1}\)$,
$\(\ref{cond2}\)$, $\(\ref{cond5}\)$ and $\(\ref{cond6}\)$
for the right-moving modes become
\begin{eqnarray}
\left [ a_n^{I}
-e^{\(\frac{\beta}{2}\hat{H}+i\pi\lambda\hat{P}\)}
\widetilde{a}_n^{\dagger\:I}\right] |0\(\theta\)\rangle
&=&0,
\label{bondop1}
\\
\left[\widetilde{a}_{n}^{I}-
e^{-\(\frac{\beta}{2}\hat{H}+i\pi\lambda\hat{P}\)}
{a}^{\dagger\: I}_{n}\right]
|0\(\theta\)\rangle
&=&0,
\\
\langle 0\(\theta\)|
\left[a^{\dagger\: I}_{n}-\widetilde{a}_{n}^{I}
e^{\(\frac{\beta}{2}\hat{H}+i\pi\lambda\hat{P}\)}\right]
&=&0,
\\
\langle 0\(\theta\)|
\left[\widetilde{a}^{\dagger\: I}_n
-a_{n}^{I}e^{-\(\frac{\beta}{2}\hat{H}+i\pi\lambda\hat{P}\)} \right]
&=&0.
\label{bondop2}
\end{eqnarray}
The above expressions $\(\ref{bondop1}\)$ and $\(\ref{bondop2}\)$,
for example, can be extended to \cite{ume1}
\begin{eqnarray}
\left[A
-e^{\(\frac{\beta}{2}\hat{H}+i\pi\lambda\hat{P}\)}
\widetilde{A}^{\dagger}\right]|0\(\theta\)\rangle&=&0,
\label{bondA}
\\
\langle 0\(\theta\)|\left[\widetilde{A}^{\dagger}-A
e^{-\(\frac{\beta}{2}\hat{H}+i\pi\lambda\hat{P}\)}\right]
&=&0,
\label{bondB}
\end{eqnarray}
where A stands for any sum of normal ordered products of
$a^{I}_{n}$ and $a_{n}^{\dagger \: I}$.

Now, we are able to show that the thermal state
conditions $\(\ref{cond1}\)$-$\(\ref{cond8}\)$,
that are in fact related with the identifications
$\(\ref{ident}\)$, give us a generalized
KMS condition for closed string theory.
In usual quantum field theory, defining two
operators $A\(t\)$ and $B\(t'\)$, the KMS
condition reads
\begin{equation}
\left\langle 0 (\b ) \left|A\(t\)B\(t'\)
\right| 0 (\b )\right\rangle
= \left\langle 0 (\b ) \left| B\(t'\)A\(t+i\b\)
\right| 0 (\b )\right\rangle.
\label{KMS}
\end{equation}
Let us derive this condition when A and B stand for closed
string right-moving world-sheet fields.
Since $\hat{H}$ and $\hat{P}$ generate time and $\sigma$
translation, equations $\(\ref{bondA}\)$ and
$\(\ref{bondB}\)$ can be written as
\begin{eqnarray}
\left[A\(\tau+i\beta,\sigma-2\pi\lambda\)
-\widetilde{A}^{\dagger}\(\tilde{\tau}-i\frac{\b}{2},
\tilde{\sigma}+\pi\lambda\)\right]\left|0\(\beta\)\right\rangle
&=&0,
\nonumber
\\
\langle 0\(\beta\)|
\left[\widetilde{A}^{\dagger}\(\tilde{\tau}-i\frac{\beta}{2},
\tilde{\sigma}+\pi\lambda\)
-A\(\tau,\sigma\)\right]
&=&0.
\end{eqnarray}
By using the above expressions, one can derive the following result
\begin{eqnarray}
\left\langle 0 (\b ) \left|A\(\tau,\sigma\)B\(\tau',\sigma'\)
\right| 0 (\b )\right\rangle
&=& \left\langle 0 (\b ) \left|
\widetilde{A}^{\dagger}\(\tilde{\tau}-i\frac{\b}{2},
\tilde{\sigma}+\lambda \pi\)B\(\tau',\sigma'\)
\right| 0 (\b )\right\rangle
\nonumber
\\
&=&\left\langle 0 (\b ) \left|B\(\tau',\sigma'\)
\widetilde{A}^{\dagger}\(\tilde{\tau}-i\frac{\b}{2},\tilde{\sigma}
+\lambda\pi\)\right| 0 (\b )\right\rangle
\nonumber
\\
&=&\left\langle 0 (\b ) \left|B\(\tau',\sigma'\)
A\(\tau+i\b,\sigma - 2\pi\lambda\)
\right| 0 (\b )\right\rangle.
\label{KMSR}
\end{eqnarray}
The same procedure applied to the bar operators gives
\begin{equation}
\left\langle 0 (\b ) \left|\bar{A}\(\tau,\sigma\)\bar{B}\(\tau',\sigma'\)
\right| 0 (\b )\right\rangle
=\left\langle 0 (\b ) \left|\bar{B}\(\tau',\sigma'\)
\bar{A}\(\tau+i\b,\sigma + 2\pi\lambda\)
\right| 0 (\b )\right\rangle.
\label{KMSL}
\end{equation}
The above results, $\(\ref{KMSR}\)$ and $\(\ref{KMSL}\)$, are a
generalization of the KMS condition, when the torus topology is
taken into account. It is a direct consequence of the thermal
states conditions.

\section{Conclusions}

In this work the Thermo Field Dynamics (TFD) formalism was
used to treat closed strings at thermal equilibrium.
The main characteristic of TFD is the construction of a thermal
Fock space and thermal operators. We give a topological
interpretation for the thermal vacuum, relating this approach
with the Imaginary Time formalism.

The TFD algorithm starts by defining an auxiliary Hilbert
space and a thermal vacuum is constructed by means of a Bogoliubov
transformation, that entangles the elements of the two Hilbert
spaces. We show that the auxiliary Hilbert space is related with a
string that propagates backwards in Euclidian time. The effect of the
thermal Bogoliubov transformation, in the equilibrium situation,
is to identify the two strings in order to make a torus with
moduli space parameters defined by $\t= \lambda +
i\frac{\b}{2\pi}$, where $\lambda$ is a Lagrange multiplier
fixing the closed string's $S^1$ isometry. We show that these
boundary equations are in fact the thermal state conditions and
the thermal vacuum is the boundary state solution that comes from
the identification of the two strings. With this we clarify why
the TFD computations carried out in our previous works
\cite{nos,nos2,nos3} reproduce the results coming from the
Imaginary Time formalism for free closed strings, where the theory
is defined on a torus. The interpretation for the torus in TFD
presented here agrees with that given by Laflamme \cite{laf},
where the original and tilde fields live in different surfaces.
Also, as a consequence of the thermal state conditions, we provided
a generalization of the KMS boundary condition that holds when the
torus topology is taken into account.

There are many possible extensions of this work. We can try to 
use the techniques developed in \cite{Gaume} to further use 
TFD to study string theory at finite temperature at higher genus.
The way how TFD works under these
situation consists by itself in an interesting point for
investigation. Also, as this is a real time formalism, it can be
used in a more involved situation, taking into account for
example, time dependent geometries. In fact, due to the evolution
of the particle distribution, a real time formalism seems the
appropriated one to be used in theories containing gravity
\cite{mathur}. Furthermore the possibility to go out of thermal
equilibrium can be very useful for applications in string
cosmology.

\section*{Acknowledgements}
We would like to thank D. Z. Marchioro, M. Ruzzi, Brenno Carlini Vallilo
and I. V. Vancea for useful discussions. M. C. B. A. was partially
supported by the CNPq grant 302019/2003-0, A. L. G. and D. L. N.
are supported by a FAPESP post-doc fellowship.


\begin{thebibliography}{99}

\bibitem{mat1}
T. Matsubara, Prog. Theor. Phys. {\bf 14} (1955) 351.

\bibitem{le1}
A. L. Fetter and J. D. Walecka,
{\it Quantum Theory of Many-Particles Systems}
(McGraw-Hill, New York, 1971).

\bibitem{adas}
A. Das, {\it Finite Temperature Field Theory}
(W. Scientific, Singapore, 1977).

\bibitem{laz}
l. Castellani, Riccardo D'Auria and Pietro Fr\'{e},
{\it Supergravity and Superstrings - A Geometric
Perspective}, Vol. 3
(World Scientific Publishing Co. Pte. Ltd, Singapore, 1991).

\bibitem{haag}
R. Haag,
{\it Local Quantum Physics: Fields, Particles,
Algebras}
(Springer-Verlag, New York, 1992).

\bibitem{emch}
G. G. Emch,
{\it Alebraic Methods in Statiscal and Quantum Field Theory}
(John Wiley, New York, 1972).

\bibitem{ume2}
Y. Takahashi and H. Umezawa,
Coll. Phenomena {\bf 2} (1975) 55
(Reprinted in Int. J. Mod. Phys. {\bf 10} (1996) 1755).

\bibitem{ume4}
H. Umezawa, H. Matsumoto, M. Tachiki,
{\it \ Thermofield Dynamics and Condensed States}
(North-Holland, Amsterdan, 1982).

\bibitem{rev2}
A. Mann, M. Revzen, H. Umezawa and Y. Yamanaka,
Phys. Lett. A {\bf 140} (1989) 475.

\bibitem{ume1}
H. Umezawa,
{\it Advanced Field Theory: Micro, Macro and Thermal Physics }
(AIP, New York, 1993).

\bibitem{kha2}
A. E. Santana, F. C. Khanna, Phys. Lett. A {\bf 203} (1995) 68.

\bibitem{kha3}
A. E. Santana, F. C. Khanna, H. Chu and C. Chang,
Annals Phys. {\bf 249} (1996) 481.

\bibitem{kob1}
R. Kobes, Phys. Rev. D {\bf 42} (1990) 562.

\bibitem{leb1}
M. Le Bellac,
{\it Thermal Field Theory}
(Cambridge University Press, Cambridge, 1996).

\bibitem{nos}
Daniel~L.~Nedel, M.~C.~B.~Abdalla and A.~L.~Gadelha,
Phys. Lett. B {\bf 598} (2004) 121.

\bibitem{nos2}
M.~C.~Abdalla, A.~L.~Gadelha and Daniel~L.~Nedel,
Phys. Lett. A {\bf 344} (2005) 123.

\bibitem{nos3}
M.~C.~Abdalla, A.~L.~Gadelha and Daniel~L.~Nedel, in Proceedings of
``Fourth International Winter Conference on Mathematical Methods in Physics'',
PoS(WC2004)032. http://pos.sissa.it/.

\bibitem{FNA1}
H.~Fujisaki and K.~Nakagawa,
Europhys.\ Lett.\  {\bf 35} (1996) 493.

\bibitem{FNA2}
K.~Nakagawa and H.~Fujisaki,
Europhys.\ Lett.\  {\bf 28} (1994) 1.

\bibitem{YLE3}
Y.~Leblanc,
Phys.\ Rev.\ D {\bf 38} (1988) 3078.

\bibitem{FNA3}
H.~Fujisaki, S.~Sano and K.~Nakagawa,
Nuovo Cim.\ A {\bf 110} (1997) 161.

\bibitem{YLE1}
Y.~Leblanc,
Phys.\ Rev.\ D {\bf 36} (1987) 1780.

\bibitem{IVV}
I.~V.~Vancea,
Phys.\ Lett.\ B {\bf 487} (2000) 175.

\bibitem{AGV2}
M.~C.~B.~Abdalla, A.~L.~Gadelha and I.~V.~Vancea,
Phys.\ Rev.\ D {\bf 64} (2001) 086005.

\bibitem{AEG}
M.~C.~B.~Abdalla, E.~L.~Graca and I.~V.~Vancea,
Phys.\ Lett.\ B {\bf 536} (2002) 144.

\bibitem{AGV3}
M.~C.~B.~Abdalla, A.~L.~Gadelha and I.~V.~Vancea,
Phys.\ Rev.\ D {\bf 66} (2002) 065005.

\bibitem{AGV4}
M.~C.~B.~Abdalla, A.~L.~Gadelha and I.~V.~Vancea,
Int.\ J.\ Mod.\ Phys.\ A {\bf 18} (2003) 2109.

\bibitem{AGV5}
M.~C.~B.~Abdalla, A.~L.~Gadelha and I.~V.~Vancea,
arXiv:hep-th/0308114.

\bibitem{kha5}
A. E. Santana, A. Matos Neto, J. D. M. Vianna, F. C. Khanna,
Physica A {\bf 280} (2000) 405.

\bibitem{ademir}
J.~C.~da Silva, F.~C.~Khanna, A.~Matos Neto and A.~E.~Santana,
Phys. Rev. A {\bf 66} (2002) 052101.

\bibitem{OJI}
I.~Ojima,
Annals Phys.\  {\bf 137} (1981) 1.

\bibitem{laf}
R.~Laflamme,
Nucl.\ Phys.\ B {\bf 324} (1989) 233.

\bibitem{Gaume}
L.~Alvarez-Gaume, C.~Gomez, G.~W.~Moore and C.~Vafa,
Nucl.\ Phys.\ B {\bf 303} (1988) 455.

\bibitem{mathur} Samir D. Mathur, arXiv:hep-th/9306090.
\end{thebibliography}
\end{document}